\documentclass[12pt]{article}
\usepackage{overcite}
\title{ {\bf
Integrable and superintegrable systems with spin }}
\author{\vspace{1cm}\\
         {\bf Pavel Winternitz}
         \thanks{E-mail address:
        wintern@crm.umontreal.ca}
         {\,\,and \bf \.{I}smet Yurdu\c{s}en}
        \thanks{E-mail address:
       yurdusen@crm.umontreal.ca} \\Centre de Recherches Math\'{e}matiques, Universit\'{e} de Montr\'{e}al,\\ CP 6128, Succ. Centre-Ville, Montr\'{e}al, Quebec H3C 3J7, Canada}

\date{\today}
\newpage
\begin{document}
\setlength{\baselineskip}{24pt} 
\maketitle
\setlength{\baselineskip}{5mm}
\begin{abstract}
A system of two particles with spin $s=0$ and $s=\frac{1}{2}$ respectively, moving in a plane is considered. It is shown that such a system with a nontrivial spin-orbit interaction can allow an $8$ dimensional Lie algebra of first-order integrals of motion. The Pauli equation is solved in this superintegrable case and reduced to a system of ordinary differential equations when only one first-order integral exists.
\end{abstract}

PACS numbers: 02.30.Ik, 03.65.-w, 11.30.-j, 25.80.Dj

\section{INTRODUCTION}\label{intro}
A Hamiltonian system with $n$ degrees of freedom in classical mechanics is called integrable if it allows $n$ functionally independent integrals of motion $\{X_1,\ldots,X_n\}$. These integrals must be well-defined functions on phase space and be in involution. The Hamiltonian $H$ belongs to this set of $n$ integrals. A ``superintegrable system" is one that allows some additional integrals of motion, $\{Y_1,\ldots, Y_k\}$ such that the set $\{X_1,\ldots,X_n,Y_1,\ldots,Y_k\}$ is functionally independent. The integrals $\{Y_1,\ldots,Y_k\}$ are not necessarily in involution among each other, nor with the $X_i$. A system is maximally superintegrable if we have $k=n-1$, minimally superintegrable for $k=1$.

The concepts of integrability and superintegrability are also introduced in quantum mechanics. The only difference is that the integrals of motion are now well-defined linear quantum mechanical operators, assumed to be algebraically independent \cite{Fock, Bargmann, Tauch, Moshinsky, Fris, Winternitz, Makarov, Evans.a, Evans.b, Kalnins, Sheftel, Tempesta}.

The best known superintegrable systems are the Kepler, or Coulomb system~\cite{Fock, Bargmann} and the harmonic oscillator \cite{Tauch, Moshinsky}. They are characterized by the fact that all finite classical trajectories in these systems are periodic. In quantum mechanics these systems are exactly solvable, i.e. their bound state energy levels can be calculated algebraically and their wave function expressed in terms of polynomials.

The above properties are shared by all other known maximally superintegrable systems (see e.g. \cite{Tempesta}).

A systematic search for superintegrable systems and their properties was started quite some time ago \cite{Fris, Winternitz, Makarov, Evans.a, Evans.b}. Originally the approach concentrated on Hamiltonians of the type
\begin{equation}
H=-\frac{1}{2} \Delta + V(\vec{r}) \label{generalhamiltonian}
\end{equation}
in $2$- and $3$- dimensional Euclidean spaces with the restriction that all integrals of motion should be first- or second-order polynomials in the momenta. More recently the study of superintegrable systems with second-order integrals of motion was extended  to curved spaces and also higher-dimensional ones (see \cite{Kalnins} for some recent results and an extensive list of references). 

For Hamiltonians of the type (\ref{generalhamiltonian}) with second-order integrals of motion there is a close relation between integrability and the separation of variables in the Schr\"{o}dinger and Hamilton-Jacobi equations. Typically, superintegrable systems of this type are multiseparable: they allow the separation of variables in more than one system of coordinates. 

This relationship between integrability and separability breaks down in other cases. Thus for natural Hamiltonians of the type (\ref{generalhamiltonian}) the existence of third-order integrals of motion does not lead to the separation of variables \cite{Drach, Drach2, Gravel.a, Gravel.b}. Furthermore, if we consider velocity dependent potentials (e.g. related to magnetic field),  
\begin{equation}
H=-\frac{1}{2} \Delta + V(\vec{r})+(\vec{A}, \vec{p})\,, \label{velocityhamiltonian}
\end{equation}
then quadratic integrability no longer implies the separation of variables \cite{Dorizzi, Berube, Pucacco, Tempesta2}. 

In this article we initiate the study of integrability and superintegrability in a different type of system, namely one involving particles with spin. More specifically, we consider two nonrelativistic quantum particles, moving in a plane, one with spin $\frac{1}{2}$, the other with spin $0$. In this case the Hamiltonian will be a matrix operator, acting on two component spinors and we can decompose it in terms of the identity matrix $I$ and the Pauli matrices $\sigma_i$ $(i=1,2,3)$.

From the physical point of view the most interesting Hamiltonian to consider would be 
\begin{eqnarray}
H=-\frac{\hbar^2}{2m} \Delta + V_0(\vec{r}) + \frac{1}{2}\Big\{V_1(\vec{r}), (\vec{\sigma}, \vec{L})\Big\}
\end{eqnarray}
in the $3$- dimensional Euclidean space $E_3$ (we drop the matrix $I$ whenever this does not cause confusion). The curly bracket represents an anticommutator and $V_a(\vec{r})$, $a=0,1$ are real functions. The Hamiltonian is Hermitian and satisfies the requirements of parity and time reversal invariance. The spin-orbital interaction term $V_1(\vec{r})(\vec{\sigma}, \vec{L})$ is the standard one in quantum mechanics \cite{Landau}. 

In this paper, the first one on integrability and superintegrability for particles with nonzero spin, we restrict to the $2$- dimensional space $E_2$. The orbital angular momentum $\vec{L}$ then only has one component, $L_3$, perpendicular to the $xy$- plane $E_2$. The scalar product $(\vec{\sigma}, \vec{L})$ reduces to ${\sigma}_3 L_3$ (since $L_1$ and $L_2$ are zero). We shall set the reduced mass $m$ of the two particle system equal to $m=1$ and use units in which the Planck constant is $\hbar=1$ (we do not need to consider a classical limit here). Finally, the Hamiltonian to be considered in this article is:
\begin{equation}
H=\frac{1}{2}({p_1}^2+{p_2}^2)+V_0(x,y)+V_1(x,y){\sigma}_3 L_3 + \frac{1}{2}{\sigma}_3(L_3 V_1(x,y)) \label{hamiltonian}
\end{equation}
with 
\begin{eqnarray}
p_1=-i\partial_x, \qquad p_2=-i\partial_y, \qquad L_3=i(y\partial_x-x\partial_y), \qquad \sigma_3=\left(\begin{array}{cc} 1& 0\\ 0& -1 \end{array}\right)\,.\label{def}
\end{eqnarray}

A priori the functions $V_0(x,y)$ and $V_1(x,y)$ are arbitrary. In specific physical applications they may be related or they may both be specified. Our aim is to determine the conditions on these two functions, under which one or more integrals of motion exist.

We request that at least one first-order integral of motion should exist, namely
\begin{equation}
X=(A_0 p_1 + B_0 p_2 + \phi_0) I + (A_1 p_1 +B_1 p_2 + \phi_1)\sigma_3\,, \label{integralofmotion}
\end{equation}
where $A_{\mu}$, $B_{\mu}$ and $\phi_{\mu}$ ($\mu=0,1$) are real functions of $x$ and $y$. These functions as well as the potentials $V_0$ and $V_1$ are to be determined from the commutativity condition
\begin{equation} 
[H,X]=0\,.\label{commutator}
\end{equation}

The general formulation is set up in Section~\ref{two} where we determine $A_{\mu}$ and $B_{\mu}$ and  obtain the partial differential equations (PDE) that $\phi_0, \phi_1, V_0$ and $V_1$ must satisfy. In Section~\ref{three} we consider a special case when the Hamiltonian (\ref{hamiltonian}) allows $6$ independent nontrivial integrals of the type (\ref{integralofmotion}). They generate an $8$-dimensional symmetry group of the system. Section~\ref{four} is devoted to more general integrable Hamiltonians, allowing just $1$ first-order integral. The system of equations 
\begin{eqnarray}
H\Psi=E\Psi, \qquad \quad X\Psi=\lambda \Psi \label{eigenvalue}
\end{eqnarray}
is studied in Section~\ref{five}. We shall call the system (\ref{eigenvalue}) the Pauli system. Some conclusions and future directions are outlined in the final Section~\ref{six}.

\section{FORMULATION OF THE PROBLEM}\label{two}
In order to obtain determining equations for the coefficients $A_{\mu}$, $B_{\mu}$ and $\phi_{\mu}$, ($\mu=0,1$) in the integral (\ref{integralofmotion}) we impose the commutation relation (\ref{commutator}). The commutator will involve terms of the type $p_1^2$, $p_1^2 \sigma_3$, $p_2^2$, $p_2^2 \sigma_3$, $p_1p_2$, $p_1p_2 \sigma_3$, $p_1$, $p_1 \sigma_3$, $p_2$, $p_2 \sigma_3$, $I$ and $\sigma_3$. We have $\sigma_3^2=I$, so no higher powers of $\sigma_3$ appear. We set the coefficients of each of the above terms equal to zero. This gives us $12$ linear partial differential equations for $A_{\mu}$, $B_{\mu}$ and $\phi_{\mu}$. Those coming from the coefficients of second-order terms in the momentum imply that $A_{\mu}$ and $B_{\mu}$ are linear functions and we obtain, for any potentials $V_0$ and $V_1$ 
\begin{eqnarray}
A_{\mu}=\omega_{\mu}y+a_{\mu}, \qquad B_{\mu} = -\omega_{\mu}x + b_{\mu}\,, \label{linear}
\end{eqnarray}
where $\omega_{\mu}$, $a_{\mu}$ and $b_{\mu}$ are real constants. The coefficients of $\vec{p}$, $\vec{p}\,\sigma_3$, $I$ and $\sigma_3$ in the commutator provide an overdetermined system of six first-order PDEs for the four functions $V_0$, $V_1$, $\phi_0$ and $\phi_1$. They are
\begin{eqnarray}
\phi_{\mu, x}&=& \delta_{\mu, 1-\nu}[-b_{\nu}V_1-(\omega_{\nu}y+a_{\nu})y V_{1,x}+(\omega_{\nu}x-b_{\nu})y V_{1,y}]\,, \nonumber \\
\label{combatibility1}\\
\phi_{\mu, y}&=& \delta_{\mu, 1-\nu}[a_{\nu}V_1+(\omega_{\nu}y+a_{\nu})x V_{1,x}-(\omega_{\nu}x-b_{\nu})x V_{1,y}]\,, \nonumber
\end{eqnarray}
\begin{eqnarray}
(\omega_{\mu}y+a_{\mu})V_{0,x}+(-\omega_{\mu}x+b_{\mu})V_{0,y}=\delta_{\mu, 1-\nu}(x\phi_{\nu,y}-y\phi_{\nu,x})V_1\,, \qquad (\mu, \nu=0,1)\,.\label{combatibility2}
\end{eqnarray}

The coefficients of $I$ and $\sigma_3$ a priori involve second-order derivatives of $V_1(x,y)$. These second-order terms cancel, once equations (\ref{linear}) and (\ref{combatibility1}) are taken into account. This leads to the two first-order equations (\ref{combatibility2}).
 
Before solving this system, let us introduce ``allowed transformations" that leave the Hamiltonian (\ref{hamiltonian}) form invariant, i.e. change only the functions $V_0(x,y)$ and $V_1(x,y)$. Such transformations will be used to simplify Hamiltonians, integrals of motion and also the equations to be solved.

Allowed transformations for any potentials $V_0$ and $V_1$ are:
\begin{enumerate}
\item Rotations in the $xy$-plane.
\item Gauge transformations of the form
\begin{eqnarray}
\tilde H=U^{-1}HU, \qquad U=\left(\begin{array}{cc} e^{i\alpha}& 0\\ 0& e^{-i\alpha} \end{array}\right), \qquad \alpha=\alpha(\xi), \qquad \xi=\frac{y}{x}\,. \label{gauge}
\end{eqnarray}
The transformation of the potentials is 
\begin{eqnarray}
\tilde {V_1}=V_1 + \frac{\dot {\alpha}}{x^2}, \qquad \tilde {V_0}=V_0+(1+\frac{y^2}{x^2})(\frac{1}{2}\frac{{\dot {\alpha}}^2}{x^2}+ \dot {\alpha}V_1)\,. \label{gaugetransform}
\end{eqnarray}
\end{enumerate} 

For certain specific potentials $V_1$ further allowed transformations exist (for any $V_0$), namely simultaneous translations and gauge transformations
\begin{enumerate}
\item $V_1=\gamma=$ const \\
The allowed transformations are given by 
\begin{eqnarray}
\tilde x = x+x_0\,,  \qquad \tilde y = y+y_0\,, \qquad \alpha = \gamma (y_0 x- x_0 y)
\end{eqnarray}
and the transformation of the potentials is
\begin{eqnarray}
\tilde {V_1} = V_1=\gamma\,,  \qquad \tilde {V_0}(x,y) = V_0(x+x_0, y+y_0)-\frac{1}{2}{\gamma}^2({x_0}^2+{y_0}^2+2(xx_0+yy_0))\,.\label{transgauge1}
\end{eqnarray}
\item $V_1=V_1(x)$ \\
The allowed transformation is 
\begin{eqnarray}
\tilde x = x\,,  \qquad \tilde y = y+y_0\,, \qquad \alpha (x) = y_0 \int V_1(x)dx\,, \nonumber \\
\tilde {V_1} = V_1,  \qquad \tilde {V_0}(x,y) = V_0(x, y+y_0)-\frac{1}{2}{V_1}^2 y_0({y_0}+2y)\,.\label{transgauge2}
\end{eqnarray}
\end{enumerate} 

Let us now return to equation (\ref{combatibility1}). The compatibility conditions $\phi_{\mu, xy}=\phi_{\mu, yx}$ imply
\begin{eqnarray}
(\omega_{\mu}y+a_{\mu})(xV_{1,xx}+yV_{1, xy}+3V_{1,x})+(-\omega_{\mu}x+b_{\mu})(yV_{1, yy}+xV_{1, xy}+3V_{1,y})=0\,,\,\,\, (\mu=0,1)\,. \label{compatibilitymixed}
\end{eqnarray}

In general, (\ref{compatibilitymixed}) represents an overdetermined system of two different equations for the potentials $V_1(x,y)$ and this system can be written as 
\begin{eqnarray}
xV_{1,xx}+yV_{1, xy}+3V_{1,x}=0\,, \nonumber \\
yV_{1, yy}+xV_{1, xy}+3V_{1,y}=0\,. \label{system}
\end{eqnarray}

An exception occurs if the two equations (\ref{compatibilitymixed}) coincide. This happens if the constants figuring in equation (\ref{compatibilitymixed}) satisfy the three following equations:
\begin{eqnarray}
\omega_0a_1-\omega_1a_0=0, \qquad \omega_0b_1-\omega_1b_0=0, \qquad a_1b_0-a_0b_1=0\,.\label{determinant}
\end{eqnarray}

We shall treat the case (\ref{system}) in Section~\ref{three} below. The case when (\ref{determinant}) is satisfied and $V_1(x,y)$ satisfies only one equation (\ref{compatibilitymixed}) will be considered in Section~\ref{four}.

\section{SPIN ORBITAL INTERACTION WITH \newline KINEMATICAL INVARIANCE GROUP}\label{three}
Let us now solve equations (\ref{system}). We transform the first equation to characteristic variables, solve and substitute into the second equation. The result is 
\begin{eqnarray}
V_1(x,y)=\gamma + \frac{G(\xi)}{x^2}, \qquad \xi=\frac{y}{x}, \qquad \gamma = \mbox{const} \,.\label{pot1}
\end{eqnarray}
Comparing with equation (\ref{gaugetransform}) we see that we can annul the function $G(\xi)$ by a gauge transformation. Thus we have
\begin{equation}
V_1=\gamma\,. \label{pot2}
\end{equation}
Substituting (\ref{pot2}) into equations (\ref{combatibility1}) and (\ref{combatibility2}) we obtain
\begin{eqnarray}
V_0=\frac{1}{2}{\gamma}^2(x^2+y^2), \qquad \phi_0=-\gamma (b_1x-a_1y), \qquad \phi_1=-\gamma (b_0x-a_0y)\,.
\end{eqnarray}
The Hamiltonian thus has the form
\begin{eqnarray}
H=-\frac{1}{2}\Delta + \frac{1}{2}{\gamma}^2(x^2+y^2) + \gamma \sigma_3 L_3, \qquad \gamma \in R\,. \label{hamiltoniansuper}
\end{eqnarray}
Since $H$ does not depend on the constants $\omega_{\mu}$, $a_{\mu}$ and $b_{\mu}$ we obtain 6 independent integrals of motion, generating the symmetry group of this Hamiltonian. 

A basis for the symmetry algebra is given by the 8 operators
\begin{eqnarray}
L_{\pm}&=&i(y\partial_x-x\partial_y)(I\pm \sigma_3)\,,\nonumber \\
X_{\pm}&=&(i\partial_x \mp \gamma y )(I\pm \sigma_3)\,,\nonumber \\
Y_{\pm}&=&(i\partial_y \pm \gamma x )(I\pm \sigma_3)\,,\nonumber \\
I_{\pm}&=& I\pm \sigma_3\,. \label{symmetryalgebra}
\end{eqnarray}
The nonzero commutation relations are 
\begin{eqnarray}
[L_{\pm}, X_{\pm}]=2iY_{\pm}, \qquad [L_{\pm}, Y_{\pm}]=-2iX_{\pm}, \qquad [X_{\pm}, Y_{\pm}]={\pm}4i\gamma I_{\pm}\,. \label{algebracommutations}
\end{eqnarray} 
The symmetry algebra is thus isomorphic  to the direct sum of two central extensions of the Euclidean Lie algebra
\begin{eqnarray}
L \sim \tilde {e}_+(2)\oplus\tilde {e}_-(2)=\{L_+, X_+, Y_+, I_+\}\oplus \{L_-, X_-, Y_-, I_-\}\,.\label{directsum}
\end{eqnarray}
The Casimir operators of $L$ are
\begin{eqnarray}
C_{\pm}={X_{\pm}}^2 + {Y_{\pm}}^2 \pm 4\gamma L_{\pm}I_{\pm}\,,\qquad I_{\pm}=I\pm \sigma_3 \label{casimir}
\end{eqnarray}
and we have
\begin{eqnarray}
H=\frac{1}{8}(C_++C_-)\,.\label{hamiltoncasimir}
\end{eqnarray}

The integral of motion $X$ is a linear combination of the 8 operators (\ref{symmetryalgebra})
with arbitrary real constant coefficients. Such operators $X$ can be classified into conjugacy classes under the action of the group generated by the algebra (\ref{symmetryalgebra}). The conjugacy classes that lead to different types of solutions of the Pauli system (\ref{eigenvalue}) can be represented by 
\begin{eqnarray}
X_1=L_+ + \alpha L_-, \qquad X_2= L_+ + \alpha X_-, \qquad X_3=X_+ + \alpha X_-\,,\qquad \alpha \in R\,.\label{conjugacyclass}
\end{eqnarray}

The Hamiltonian (\ref{hamiltoniansuper}) is not only integrable, but actually ``first-order superintegrable". For particles of spin 0 first-order superintegrability occurs only for free motion. Notice that if we set the spin-orbit interaction equal to zero in equation (\ref{hamiltoniansuper}) (i.e. $\gamma = 0$), we obtain free motion.

\section{HAMILTONIANS ALLOWING ONE \newline FIRST-ORDER INTEGRAL}\label{four}
Let us now consider the case when equations (\ref{determinant}) are satisfied. The two equations (\ref{compatibilitymixed}) then coincide and the potential $V_1(x,y)$ satisfies just one second-order PDE. The equation (\ref{compatibilitymixed}) is of hyperbolic type. Its characteristic variables are 
\begin{eqnarray}
\xi=\frac{y}{x}, \qquad \eta=\frac{1}{2}\omega_1 (x^2+y^2)-b_1x+a_1y\,. \label{characteristics}
\end{eqnarray}

Here we shall just consider two interesting special cases.
\begin{description}
\item[a)] $\omega_1\ne 0,\, a_0=b_0=0,\, a_1=b_1=0$ \\
We transform equation (\ref{compatibilitymixed}) to polar coordinates and obtain
\begin{eqnarray}
\rho V_{1, \rho \theta}+2 V_{1,\theta}=0\,, \qquad \quad x=\rho \cos \theta,\quad y=\rho \sin \theta\,. \label{polar}
\end{eqnarray}
The potential $V_1$ hence is
\begin{eqnarray}
V_1=f(\rho)+\frac{1}{\rho^2}g(\theta) \,,\label{potentialonepolar}
\end{eqnarray}
where $f(\rho)$ and $g(\theta)$ are arbitrary. Comparing with equation (\ref{gaugetransform}) we see that the function $g(\theta)$ can be set equal to $g=0$ by a gauge transformation. Solving (\ref{combatibility1}) and (\ref{combatibility2}) we obtain
\begin{eqnarray}
V_0=V_0(\rho)\,, \qquad V_1&=&V_1(\rho)\,, \qquad \phi_0=\phi_1=0 \,, \label{fourpolarpot}\\
X &=&(\omega_0 + \omega_1 \sigma_3)L_3\,. \label{fourintegralmot}
\end{eqnarray}
\item[b)] $\omega_0=\omega_1=0$,\,  ${a_1}^2+{b_1}^2 \neq 0$ \\
Equation (\ref{compatibilitymixed}) in characteristic variables (\ref{characteristics}) is
\begin{equation}
V_{1, \xi \eta}+\frac{2}{\eta}V_{1, \xi}=0 \label{section4pot1}
\end{equation}
and we obtain
\begin{equation}
V_1=F_1(a_1y-b_1x) + \frac{F_2(\frac{y}{x})}{(a_1y-b_1x)^2}\,. \label{sec4resultpot1}
\end{equation}
By a gauge transformation we set $F_2=0$ and rotate in the $xy$- plane to obtain
\begin{equation}
V_1=V_1(x)\,. \label{sec4latestresultpot1}
\end{equation}
From equations (\ref{combatibility1}) and (\ref{combatibility2}) we obtain
\begin{eqnarray}
V_0=\frac{y^2}{2} {V_1}^2(x) + F(x) \,, \label{fourpotforcartesiancase}\\
\phi_0=-b_1 \int V_1(x)dx\,, \qquad \phi_1=-b_0 \int V_1(x)dx \,.
\end{eqnarray} 
Let us put $b_1=1$, $b_0\ne0$. We then have 
\begin{eqnarray}
\phi_1=b_0\phi_0\,, \qquad V_1(x)=-\phi_0^{\prime}(x)\,, \nonumber \\
V_0(x,y)=\frac{1}{2}y^2[\phi_0^{\prime}(x)]^2+F(x) \label{fourcartesianpotlast}
\end{eqnarray}
and 
\begin{eqnarray}
X=-ib_0\partial_y+\phi_0(x)+(-i\partial_y+b_0\phi_0(x))\sigma_3\,. \label{fourcartesianintlast}
\end{eqnarray}
\end{description}
We shall call the case (\ref{fourpolarpot}) the ``polar" case, (\ref{sec4latestresultpot1}), (\ref{fourpotforcartesiancase}) the ``Cartesian" one, because of the form of the operator $X$ in (\ref{fourintegralmot}) and (\ref{fourcartesianintlast}), respectively.

\section{SOLUTIONS OF THE PAULI EQUATION}\label{five}
In this section we shall analyze and solve the pair of equations (\ref{eigenvalue}) for the different superintegrable, or first-order integrable cases, found above.
\subsection*{1. The superintegrable Hamiltonian}\label{fiveone}
Let us consider the Hamiltonian (\ref{hamiltoniansuper}) with $\gamma\ne0$, i.e. a constant spin-orbital potential and a harmonic oscillator spin-independent one. The Hamiltonian commutes with the entire kinematical algebra (\ref{symmetryalgebra}). We shall choose the operator $X$ of equation (\ref{eigenvalue}) in the form of one of the different one-dimensional subalgebras shown in equation (\ref{conjugacyclass}) and consider each of the three cases separately. The potentials $V_0(x,y)$ and $V_1(x,y)$ in this case have the form (\ref{fourpolarpot}) and (\ref{fourcartesianpotlast}) simultaneously. Hence we can separate variables in polar coordinates, as well as in Cartesian ones. Moreover, we can consider a mixed case: separation in different coordinate systems for the upper and lower components of $\Psi$. 
\subsubsection*{a) Polar case}\label{fiveonea}
We introduce polar coordinates $(\rho, \theta)$ and choose the operator $X$ in the form 
\begin{eqnarray}
X=-i\left(\begin{array}{cc} a_1& 0\\ 0& a_2 \end{array}\right)\partial_{\theta}\,, \quad a_i\ne0\,.
\end{eqnarray}
The condition $X\Psi=\lambda \Psi$ provides a wave function in the form 
\begin{eqnarray}
\Psi(\rho, \theta)=\left(\begin{array}{c} F_1(\rho)\,e^{i\frac{\lambda}{a_1}\theta}\\ F_2(\rho)\,e^{i\frac{\lambda}{a_2}\theta} \end{array}\right)\,.\label{polarcaseseparationwavefunc}
\end{eqnarray}
Substituting into the Pauli equation with Hamiltonian (\ref{hamiltoniansuper}) we find that the function $F_i(\rho)$ satisfy
\begin{eqnarray}
\left\{-\frac{1}{2}\Big(\frac{{\partial}^{\,2}}{{\partial \rho}^2}+\frac{1}{\rho}\frac{\partial}{\partial \rho}-\frac{1}{\rho^2}\frac{\lambda^2}{{a_i}^2}\Big)+\frac{1}{2}\gamma^2 \rho^2\right\}F_i=(E\mp\gamma \frac{\lambda}{a_i})F_i\,,\label{polarcaseharmonic}
\end{eqnarray}
so both components satisfy radial harmonic oscillator type equations. 

The solution of equation (\ref{polarcaseharmonic}) is 
\begin{eqnarray}
F_i(\rho)=N_{n_im_i}\,e^{-\frac{\gamma}{2}\rho^2}\rho^{|m_i|}L_{n_i}^{|m_i|}(\gamma \rho^2)\,, \label{polarnormalization}
\end{eqnarray}
where $L_{n_i}^{|m_i|}(z)$ are Laguerre polynomials. The quantum number $\lambda$ satisfies
\begin{eqnarray}
\frac{\lambda}{a_1}=m_1\,, \qquad \frac{\lambda}{a_2}=m_2\,,
\end{eqnarray}
hence we must choose 
\begin{eqnarray}
\frac{a_2}{a_1}=\frac{m_1}{m_2}
\end{eqnarray}
rational. The energy satisfies 
\begin{eqnarray}
E-\gamma m_1 &=& \gamma (2n_1+|m_1|+1)\,, \nonumber \\
E+\gamma m_2 &=& \gamma (2n_2+|m_2|+1)\,,
\end{eqnarray}
so the two radial quantum numbers are constrained by 
\begin{eqnarray}
2(n_2-n_1)=m_1\left(\frac{a_1}{a_2}+1\right)-|m_1|\left(|\frac{a_1}{a_2}|-1\right)\,.
\end{eqnarray}
If we normalize to have 
\begin{eqnarray}
\int_0^\infty\int_0^{2\pi} (|\Psi_1|^2+|\Psi_2|^2)\,\rho d\rho\,d\theta=1\,,
\end{eqnarray}
we must put the normalization constants in (\ref{polarnormalization}) equal to
\begin{eqnarray}
N_{n_im_i}=\sqrt{\frac{\gamma^{|m_i|+1}}{2\pi}}\sqrt{\frac{n_i!}{(n_i+|m_i|)!}}\,.
\end{eqnarray}
\subsubsection*{b) Cartesian case}
 We choose the operator to be diagonalized in the form 
\begin{eqnarray}
X=\left(\begin{array}{cc} a_1 (i\partial_y+\gamma x)& 0\\ 0& a_2 (i\partial_y-\gamma x) \end{array}\right)\,, \quad a_i\ne0\,. 
\end{eqnarray}
The equation $X\Psi=\lambda \Psi$ implies
\begin{eqnarray}
\Psi=\left(\begin{array}{c} F_1(x)\,e^{-\frac{i}{a_1}(\lambda-a_1\gamma x)y}\\ F_2(x)\,e^{-\frac{i}{a_2}(\lambda+a_2\gamma x)y} \end{array}\right)\,.\label{cartesianseparablewavefun}
\end{eqnarray}
Substituting into the Pauli equation $H\Psi=E\Psi$ with $H$ as in equation (\ref{hamiltoniansuper}) we obtain 
\begin{eqnarray}
\ddot F_i - 4\gamma^2\left(x\mp\frac{\lambda}{2a_i\gamma}\right)^2F_i+2EF_i=0\,, \quad i=1,2 \label{cartesianharmonic}
\end{eqnarray}
and hence 
\begin{eqnarray}
F_i=N_{n_i}\,e^{-\gamma{\tilde x}_i^2}H_{n_i}(\sqrt{2\gamma} {\tilde x}_i)\,, \nonumber \\
E=2\gamma (n_1+\frac{1}{2})=2\gamma (n_2+\frac{1}{2})\,, \quad n_1=n_2=n\,, \nonumber \\
N_{n_i}=\sqrt{\frac{\sqrt{2\gamma}}{\sqrt{\pi}n!2^n}}\,, \qquad {\tilde x}_{1,2}=x\mp\frac{\lambda}{2a_i\gamma}\,,
\end{eqnarray}
where $H_n(z)$ is a Hermite polynomial.
\subsubsection*{c) Mixed case}
Let us take the operator $X$ in the form
\begin{eqnarray}
X=\left(\begin{array}{cc} -ia_1 \partial_{\theta}& 0\\ 0& a_2 (i\partial_y-\gamma x) \end{array}\right)\,, \quad a_i\ne0\,.
\end{eqnarray}
The wave function will then be 
\begin{eqnarray}
\Psi=\left(\begin{array}{c} F_1(\rho)\,e^{i\frac{\lambda}{a_1}\theta}\\ F_2(x)\,e^{-\frac{i}{a_2}(\lambda+a_2\gamma x)y} \end{array}\right)\,,
\end{eqnarray}
where $(\rho, \theta)$ are polar coordinates, $(x, y)$ Cartesian ones. The function $F_1(\rho)$ will satisfy equation (\ref{polarcaseharmonic}) with $i=1$, $F_2(x)$ equation (\ref{cartesianharmonic}) with $i=2$. We hence obtain 
\begin{eqnarray}
F_1(\rho)&=&N_{n_1m_1}\,e^{-\frac{\gamma}{2}\rho^2}\rho^{|m_1|}L_{n_1}^{|m_1|}(\gamma \rho^2)\,, \nonumber \\
F_2(x)&=&N_{n_2}\,e^{-\gamma{\tilde x}_2^2}H_{n_2}(\sqrt{2\gamma} {\tilde x}_2)\,,
\end{eqnarray}
with
\begin{eqnarray}
E=\gamma (m_1+2n_1+|m_1|+1)=2\gamma(n_2+\frac{1}{2})\,.
\end{eqnarray}
\subsection*{2. Hamiltonians with one first order integral}\label{fivetwo}
\subsubsection*{a) Polar case}
We consider the potential $V_0=V_0(\rho)$, $V_1=V_1(\rho)$ as in equation (\ref{fourpolarpot}). We write the integral 
(\ref{fourintegralmot}) in the form
\begin{eqnarray}
X=-i\left(\begin{array}{cc} a_1& 0\\ 0& a_2 \end{array}\right)\partial_{\theta}\,, \quad a_i\ne0
\end{eqnarray}
and the equation $X\Psi=\lambda \Psi$ implies
\begin{eqnarray}
\Psi(\rho, \theta)=\left(\begin{array}{c} F_1(\rho)\,e^{i\frac{\lambda}{a_1}\theta}\\ F_2(\rho)\,e^{i\frac{\lambda}{a_2}\theta} \end{array}\right)\,.\label{intgpolarsepwave}
\end{eqnarray}
Substituting into the Pauli equation, we find that the radial functions $F_1$, $F_2$ satisfy: 
\begin{eqnarray}
\left\{-\frac{1}{2}\Big(\frac{{\partial}^{\,2}}{{\partial \rho}^2}+\frac{1}{\rho}\frac{\partial}{\partial \rho}-\frac{1}{\rho^2}\frac{\lambda^2}{{a_i}^2}\Big)+V_0\pm V_1\frac{\lambda}{a_i}\right\}F_i=E F_i\,, \quad i=1,2\,.\label{generalpolarcaseharmonic}
\end{eqnarray}
For instance, choosing 
\begin{eqnarray}
V_0=\frac{\alpha}{\rho}\,, \qquad V_1=\frac{\beta}{\rho^2}\,,
\end{eqnarray}
we can solve equation (\ref{generalpolarcaseharmonic}) in terms of Coulomb wave functions.
\subsubsection*{b) Cartesian case}
Let us now consider $V_0$, $V_1$ and the integral $X$ as in (\ref{fourcartesianpotlast}), (\ref{fourcartesianintlast}). The equation $X\Psi=\lambda \Psi$ with $X$ as in (\ref{fourcartesianintlast}) implies
\begin{eqnarray}
\Psi=\left(\begin{array}{c} \chi_1(x)\,e^{i\frac{\lambda-(1+b_0)\phi_0(x)}{1+b_0}y}\\ \chi_2(x)\,e^{i\frac{\lambda-(1-b_0)\phi_0(x)}{b_0-1}y} \end{array}\right)\,.\label{intgcartesianwavefun}
\end{eqnarray}
The Pauli equation then reduces to the following two ODEs:
\begin{eqnarray}
\ddot \chi_i-\left\{\frac{(\lambda \mp(b_0\pm 1)\phi_0)^2}{(b_0\pm 1)^2}\mp 2x\phi_0^{\prime}\frac{\lambda \mp (b_0\pm 1)\phi_0}{(b_0\pm 1)}+2F(x)-2E\right\}\chi_i=0\,, \quad i=1,2\,.
\end{eqnarray}
To solve these equations, or analyze further, we would have to specify the two functions $F(x)$, $\phi_0(x)$.

\section{CONCLUSIONS}\label{six}
Let us first of all compare the problem of integrability and superintegrability for particles with spin $s=0$ and spin $s\ne 0$, in this case $s=\frac{1}{2}$. The spin zero case with a scalar potential in the two-dimensional Euclidean space $E_2$ corresponds to the Hamiltonian (\ref{generalhamiltonian}). First order superintegrability is trivial: it requires $V(x)=$ const. and corresponds to free motion. Superintegrability with one first-order and one second-order integral occurs for the potentials $V=\alpha r^2$, $\alpha r^{-1}$, $\alpha x$ and $\alpha x^{-2}$ \cite{Sheftel}. Quadratic superintegrability leads to $4$ families of potentials \cite{Fris, Winternitz}, each depending on $3$ significant constants and allowing the separation of variables in at least two coordinate systems. 

By contrast for $s=\frac{1}{2}$ first-order superintegrability leads to a nontrivial system, namely the Hamiltonian (\ref{hamiltoniansuper}) with the symmetry algebra (\ref{symmetryalgebra}). The Hamiltonian allows the separation of variables in polar coordinates (see (\ref{polarcaseseparationwavefunc})) and ``R-separation" in Cartesian ones (see (\ref{cartesianseparablewavefun})). Indeed, in equation (\ref{cartesianseparablewavefun}) there is a term involving the product $xy$, that does not depend on the separation constant $\lambda$. The same is true for Hamiltonians allowing just one first-order integral. In the polar case we have separation (see (\ref{intgpolarsepwave})), in the Cartesian one R-separation (\ref{intgcartesianwavefun}). 

The next step in this research program will be to look for integrable and superintegrable systems with spin in Euclidean $3$-space. This would provide a realistic and solvable model for pion-nucleon and possibly nucleon-nucleon interactions.

\section*{ACKNOWLEDGMENTS}
The research of P. W. was partly supported by a grant from NSERC of Canada. Part of this work was done when \.{I}. Y. was supported by the Scientific and Technical Research Council of Turkey (T\"{U}B\.{I}TAK) in the framework of NATO-B1 program. \.{I}. Y. also acknowledges a postdoctoral fellowship awarded by the Laboratory of Mathematical Physics of the CRM, Universit\'{e} de Montr\'{e}al.

\makeatletter
\renewcommand{\@biblabel}[1]{$^{#1}$}
\makeatother

\end{document}